\documentclass[aps,pre,twocolumn,superscriptaddress,showpacs,showkeys,longbibliography]{revtex4-1}
\usepackage[english]{babel}
\makeatletter
\def\bbl@set@language#1{%
  \edef\languagename{%
    \ifnum\escapechar=\expandafter`\string#1\@empty
    \else\string#1\@empty\fi}%
  \@ifundefined{babel@language@alias@\languagename}{}{%
    \edef\languagename{\@nameuse{babel@language@alias@\languagename}}%
  }%
  \select@language{\languagename}%
  \expandafter\ifx\csname date\languagename\endcsname\relax\else
    \if@filesw
      \protected@write\@auxout{}{\string\select@language{\languagename}}%
      \bbl@for\bbl@tempa\BabelContentsFiles{%
        \addtocontents{\bbl@tempa}{\xstring\select@language{\languagename}}}%
      \bbl@usehooks{write}{}%
    \fi
  \fi}
\newcommand{\DeclareLanguageAlias}[2]{%
  \global\@namedef{babel@language@alias@#1}{#2}%
}
\makeatother
\DeclareLanguageAlias{en}{english}
\usepackage{wrapfig,chngcntr}
\usepackage{amsmath, amsfonts, amssymb, graphicx, float}
\usepackage{amsbsy,amsthm}
\usepackage{algorithm,algorithmic}
\usepackage{mathrsfs,multirow}
\usepackage{bm,comment}
\usepackage{verbatim}
\usepackage[caption=false]{subfig}
\usepackage{algorithm}
\usepackage{url,color}
\usepackage{latexsym}

\usepackage[draft]{changes}
\usepackage{hyperref}

\begin{document}

\title{Fast consensus clustering in complex networks}

\author{Aditya Tandon}
\affiliation{School of Informatics, Computing and Engineering, Indiana University}

\author{Aiiad Albeshri}
\affiliation{Department of Computer Science, Faculty of Computing and Information Technology
King Abdulaziz University, Jeddah 21589, Kingdom of Saudi Arabia}

\author{Vijey Thayananthan}
\affiliation{Department of Computer Science, Faculty of Computing and Information Technology
King Abdulaziz University, Jeddah 21589, Kingdom of Saudi Arabia}

\author{Wadee Alhalabi}
\affiliation{Department of Computer Science, Faculty of Computing and Information Technology
King Abdulaziz University, Jeddah 21589, Kingdom of Saudi Arabia}

\author{Santo Fortunato}
\affiliation{Indiana University Network Science Institute (IUNI)}
\affiliation{School of Informatics, Computing and Engineering, Indiana University}

\begin{abstract}
Algorithms for community detection are usually stochastic, leading to different partitions for different choices of random seeds. Consensus clustering has proven to be an effective technique to derive more stable and accurate partitions than the ones obtained by the direct application of the algorithm. However, the procedure requires the calculation of the consensus matrix, which can be quite dense if (some of) the clusters of the input partitions are large. Consequently, the complexity can get dangerously close to quadratic, which makes the technique inapplicable on large graphs. Here we present a fast variant of consensus clustering, which calculates the consensus matrix only on the links of the original graph and on a comparable number of additional node pairs, suitably chosen. This brings the complexity down to linear, while the performance remains comparable as the full technique. Therefore, our fast consensus clustering procedure can be applied on networks with millions of nodes and links.

\end{abstract}
\maketitle
%
%
\section{Introduction}

Detecting communities in networks is a fundamental part of network analysis~\cite{porter09,fortunato10,fortunato16}. It is an unsupervised classification problem and as such it is ill-defined: different techniques typically find different partitions on the same network, unless the latter has a clear-cut community structure, with very dense groups which are very loosely connected to each other. 

Another important problem of the field is that community detection algorithms are {\textit{noisy}}. Since many algorithms require the use of random numbers, for initialization, optimization,  tie-breaks, etc., different random seeds may lead to different outcomes, even when the same method is applied on the same network. Consensus clustering is an effective technique to decrease the noise induced by the stochasticity of methods. The procedure consists in ``averaging" over a set of input partitions: the result, called \textit{median} or \textit{consensus} partition is usually more robust and accurate than the input partitions~\cite{lancichinetti12}.
Consensus clustering is now regularly used in the network science community~\cite{ronhovde09,sporns13,bassett13,tagarelli17,jeub18,poulin18b}. It requires the computation of a \textit{consensus matrix}, expressing how often any pair of nodes is found in the same cluster in the input partitions. The calculation concerns only the pairs of nodes that happen to be co-clustered at least once. This means that all pairs of nodes of any cluster of the input partitions will give non-vanishing entries in the consensus matrix. In particular, if the largest cluster in any input partition has a non-negligible size with respect to the whole network, the consensus matrix can become quite dense and both the space and the time complexity of the calculation could get close to $O(n^2)$, where $n$ is the number of nodes of the graph. Such high complexity limits the applicability of the procedure to networks which are not too large.

In this paper we propose a fast consensus clustering technique, which is applicable to large networks. The basic idea is to compute the consensus matrix for a very small sets of properly selected pairs of nodes. This way the matrix will be sparse all along the calculation, which can reach linear complexity, if the clustering technique used to detect the communities has itself linear complexity. If instead we have reasons to prefer to use a slower clustering algorithm its complexity will be dominating compared to the other steps of the fast consensus clustering routine developed here.

\section{The method}
\label{sec:method}

Before we get into the details of our technique, we briefly summarize the consensus clustering method exposed in Ref.~\cite{lancichinetti12}, which will be our reference and will be called \textit{LF consensus} throughout. We start from a graph $G$, with $n$ nodes and $m$ links, and a clustering algorithm of our choice. We apply the algorithm $n_p$ times on $G$, with different random seeds, obtaining $n_p$ partitions, which are the input partitions of the method.
The \textit{consensus matrix} $\textbf{D}$ is an $n \times n$ matrix. Each entry $\textbf{D}_{ij}$ corresponds to the fraction of input partitions in which nodes $i$ and $j$ belong to the same cluster. For instance, if nodes 3 and 8 are put in the same cluster in 10 out of 20 input partitions, $D_{38}=D_{83}=10/20=1/2$. By construction, the consensus matrix is then a weighted matrix, and its entries are real numbers between zero and one. Once the consensus matrix has been constructed, the original community detection algorithm is applied on it $n_p$ times.
Since all pairs of nodes in the same cluster in any partition yield a contribution, the consensus matrix can be very dense, even if the initial network is sparse. This could be a problem when we apply the clustering algorithm on it, for two reasons: 1) many algorithms have troubles to detect communities on dense matrices; 2) the computational complexity of the procedure could get very high. Therefore, before the clustering algorithm is applied, $\textbf{D}$ is filtered, in that all entries below a certain threshold $\tau$ are set to zero. Then a new consensus matrix $\textbf{D}^\prime$ is built from the new partitions, and so on. The process is repeated until all partitions are identical, i.e., the consensus matrix has only entries equal to zero and one. \\
This procedure can improve considerably the accuracy of the result, with respect to the set of initial partitions, but the construction of the consensus matrix has a worst time and space complexity of $O(n^2)$, irrespective of the sparsity of the original network, which makes the technique prohibitively expensive on large networks.  

To save both space and time, the consensus matrix should be computed only for a small subset of all eligible node pairs. The natural candidates are the pairs of nodes which are connected to each other~\cite{poulin18b}. This choice proves to be a valid one. However, there are cases in which the gain in the accuracy of the consensus partitions is very modest if we consider only the pairs of neighboring nodes. Therefore we added some additional pairs, chosen such that they close triads with the links of the original graph, provided the weights of those links in the consensus matrix are sufficiently large. This leads to the following routine:  

\begin{enumerate}
    \item {\textbf{Derive the input partitions}}. Apply the clustering algorithm on the network at study $n_p$ times, to obtain the input partitions. In all the tests shown here we took $n_p=20$.
    \item {\textbf{Construct the consensus matrix}}. Compute only the elements $\textbf{D}_{ij}$, where $i$ and $j$ are neighbors in $G$.
    \item {\textbf{Thresholding}}. Set to zero all elements of $\textbf{D}$ below a threshold value $\tau$. This removes weak links and speeds up convergence. By doing so, some nodes might get disconnected from the graph corresponding to the matrix. If a node gets disconnected, keep it attached to the rest of the graph by preserving the link with the highest weight. This way the graph is connected at all times. 
    \item {\textbf{Triadic closure}}. Select $m$ random nodes. For each node select at random a pair of neighboring nodes $j$ and $k$. If the entry $\textbf{D}_{jk}=0$, then we set it equal to the fraction of partitions in which nodes $j$ and $k$ co-occur in the same cluster. 
    \item Apply the clustering algorithm on the new weighted graph $\textbf{D}$ repeatedly to get $n_p$ new partitions. 
    \item Repeat steps 2-5 until convergence.
\end{enumerate}

Convergence is reached when less than $2\%$ of all non-zero entries of $\textbf{D}$ have weights smaller than one (a weight of one implies that the two nodes co-occur in the same cluster in all input partitions). The output is the matrix $\textbf{D}_{\textrm{out}}$. 
We then apply the community detection algorithm on $\textbf{D}_{\textrm{out}}$ to get the final set of partitions, which represent the output of our technique.
The $2\%$ threshold is suggested by our numerical experiments: for $1\%$, which was our initial choice, the number of steps required until convergence sensibly increases. However, we have verified that results are stable for thresholds at least up to $10\%$, so the actual value is not important, provided it is not too low. 

The main difference from the procedure of 
Ref.~\cite{lancichinetti12} is the fact that we compute up to $2m$ elements of $\textbf{D}$, which is then very sparse. The advantage is that the calculation of $\textbf{D}$ has space and time complexity $O(m)$, which is much lower than the $O(n^2)$ of the full method when the graph at hand is sparse ($m \propto n$), as it usually happens for real networks. On the downside,
the partitions obtained by running a clustering algorithm on $\textbf{D}$ are a bit more noisy than in the full procedure and it is unlikely to converge to a set of identical partitions in the end. This is why we accept to stop when $\textbf{D}_{\textrm{out}}$ has but a modest proportion of entries different from zero and one. However, with our criterion convergence is usually reached within a handful of iterations, and the output partitions turn out to be of superior quality than the set of input partitions, for all community detection methods we have used in our experiments. 

For our calculations we used Python implementations of the algorithms from the \href{https://networkx.github.io/}{{\tt Networkx}}~\cite{nxref} and \href{http://igraph.org/}{{\tt igraph}}~\cite{igraphref} libraries. The software to perform our fast consensus clustering procedure can be found on the following website: 
\url{http://github.com/adityat/fastconsensus}.

\section{Results}
\label{sec:results}

For our tests we used artificial benchmark graphs with built-in community structure. Specifically, we adopted the LFR benchmark graphs, which
have become a standard in the evaluation of the performance of
clustering algorithms~\cite{lancichinetti08,lancichinetti09c}. LFR graphs are characterized by power law
distributions of node degree and community size, features that frequently occur in real world networks.
The \textit{mixing parameter} $\mu$ is the ratio between the external degree of a node with respect to its community (i.e., the number of links joining the node to its neighbors outside its community) and its total degree. So, when $\mu$ is close to zero, the nodes have most of their neighbors within their communities, which are then well separated from each other and easily detectable. The larger $\mu$ the fuzzier the communities and the more difficult it gets to identify them.
\begin{figure}
\centering
 \includegraphics[width=\columnwidth]{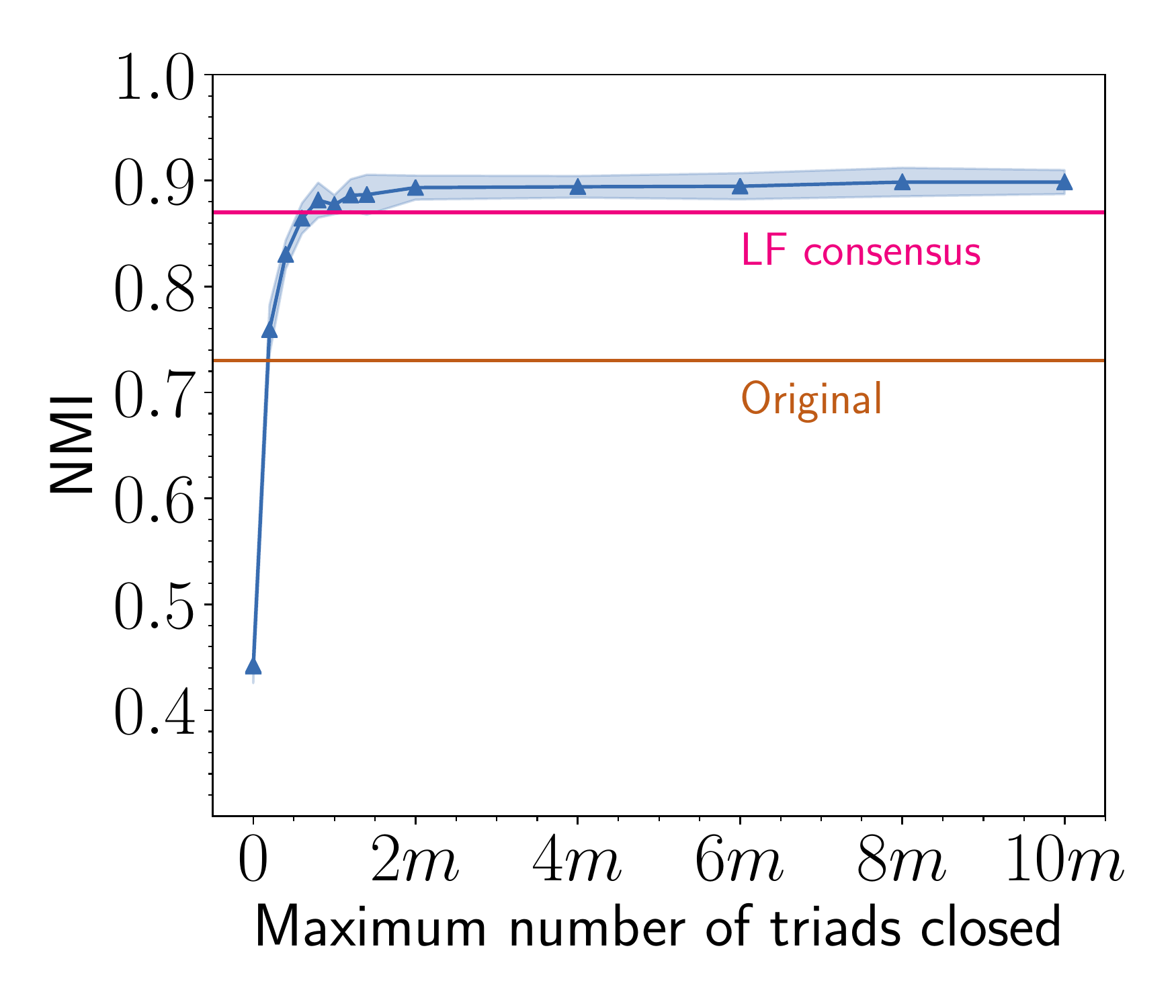}
\caption{Importance of triadic closure in the construction of a sparse consensus matrix. The accuracy of the resulting partition rapidly increases as we compute additional elements of the consensus matrix, by randomly closing triads formed by the (weighted) links of the initial matrix, after thresholding.
Closing a modest number of triads suffices to outperform the direct application of the clustering method (here the Louvain algorithm), whose accuracy is indicated by the horizontal yellow line. The accuracy quickly climbs up to that of \textit{LF consensus} (magenta line) and reaches about $0.87$ for about $m$ closed triads, remaining remarkably stable thereafter. The thickness of the blue line indicates the standard error of the mean. For the LFR benchmark graph used in this test we chose the following parameters: number of nodes $n = 10,000$, mixing parameter $\mu = 0.75$, degree exponent $\tau_1 = 2$, community size exponent $\tau_2 = 3$, average degree $k_{avg} = 20$, maximum degree $k_{max} = 50$, minimum community size $c_{min} = 10$, maximum community size $c_{max} = 100$. }
\label{fig:triad_closure}
\end{figure}
The performance of a method on the LFR networks will be estimated by computing the \textit{normalized mutual information} (NMI) between the built-in partition of the graph and the one detected by the clustering algorithm, as a function of $\mu$. We used the modified version of the NMI introduced by Lancichinetti, Fortunato and Kert\'esz~\cite{lancichinetti09}, to make the results comparable with those of Ref.~\cite{lancichinetti12}. For each $\mu$-value we created $20$ benchmark graphs (unless specified otherwise) and averaged the corresponding NMI-scores among them. 

We will show the results obtained by integrating our consensus technique with the following three clustering algorithms:

\begin{itemize} 
\item 
\textit{Fast greedy modularity optimization}. It is a technique developed
by Clauset, Newman and Moore (CNM)~\cite{clauset04}, that performs a quick maximization of the
modularity by Newman and Girvan~\cite{newman04b}. 
\item \textit{Louvain method}, by Blondel et al.~\cite{blondel08}. The goal is still the optimization of modularity, by means of a hierarchical approach. We will be using the first (bottom) level of the hierarchy generated by the method, the one with the smallest communities. This partition gives an excellent performance on LFR benchmark graphs~\cite{lancichinetti09c}. The actual outcome of the procedure, which corresponds to the top level partition and the largest modularity, is known to be poorly correlated with the built-in partitions of the benchmark~\cite{fortunato16}, mostly because of the resolution limit of modularity~\cite{fortunato07}.
\item \textit{Label Propagation Method (LPM)} by Raghavan et al.~\cite{raghavan07}. This method simulates the spreading of labels based on the simple rule that at each iteration a given
vertex takes the most frequent label in its neighborhood. 
\end{itemize}

We also run tests with the well known Infomap algorithm~\cite{rosvall08}, but this method has a great performance on LFR benchmark graphs~\cite{lancichinetti09c}, so the consensus procedure can lead just to a modest improvement and we do not show it here. We have however verified that our technique delivers higher quality results than the input partitions for Infomap as well.

Before showing the performance of our method we would like to discuss two issues.
\begin{figure}
\centering
 \includegraphics[width=\columnwidth]{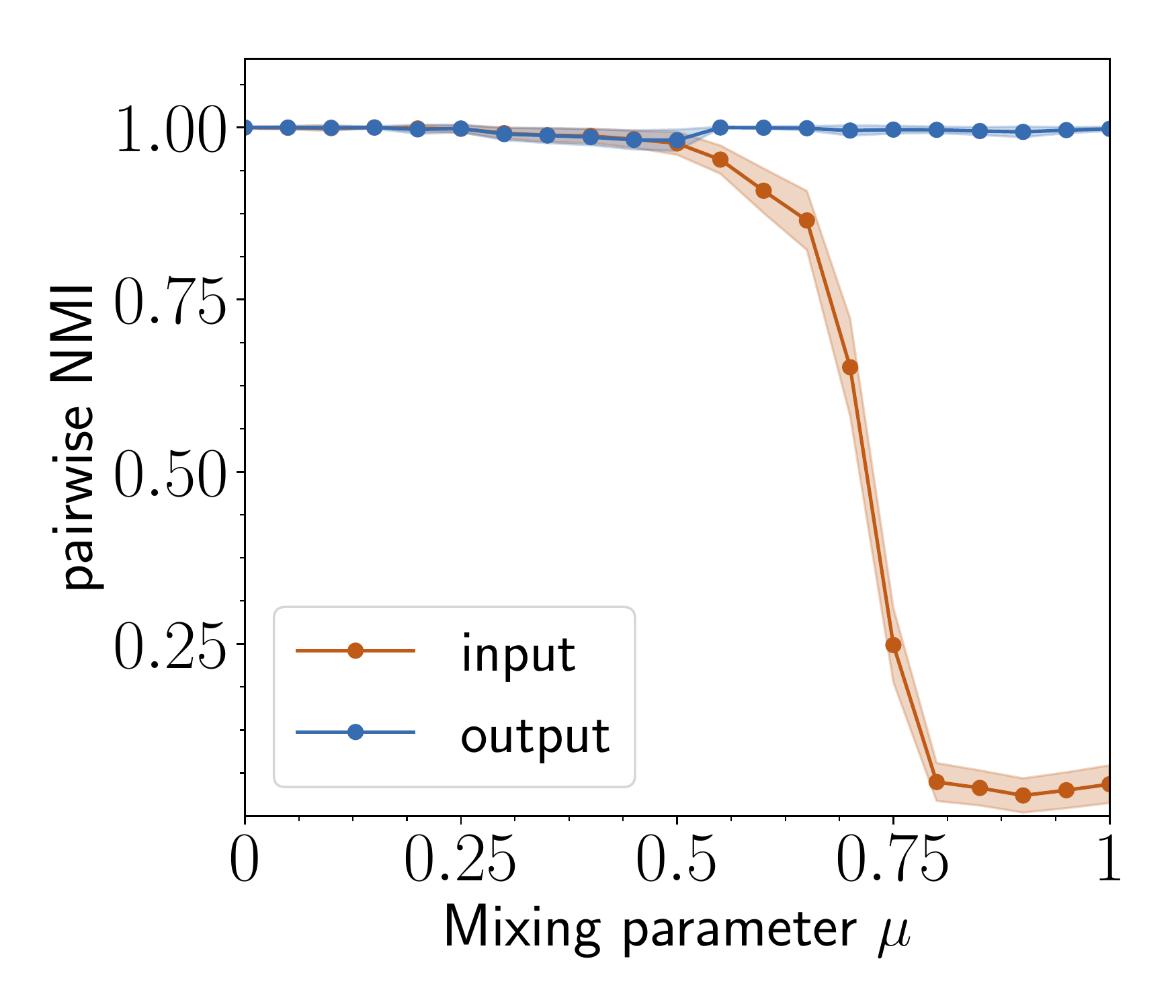}
\caption{Stability of consensus clustering partitions. Average value of the NMI for all pairs of partitions returned by our technique, compared with the corresponding average score for the input partitions.
The thickness of the lines indicate the standard error of the average.
The test uses the Louvain algorithm on LFR benchmarks. When communities are easy to find (low to intermediate $\mu$-values), partitions are essentially identical in both cases and the pairwise NMI is equal to one. Interestingly, for our consensus clustering procedure the resulting partitions are very similar even when communities are harder to find, whereas the input partitions become less and less similar to each other. The parameters used to generate the LFR benchmark graphs are: number of nodes $n=10,000$, degree exponent $\tau_1 = 2$, community size exponent $\tau_2 = 3$, average degree $k_{avg} = 20$, maximum degree $k_{max} = 50$, minimum community size $c_{min} = 10$, maximum community size $c_{max} = 100$.}
\label{fig:stability}
\end{figure}
\begin{figure*}
\centering
 \includegraphics[width = 0.98\textwidth]{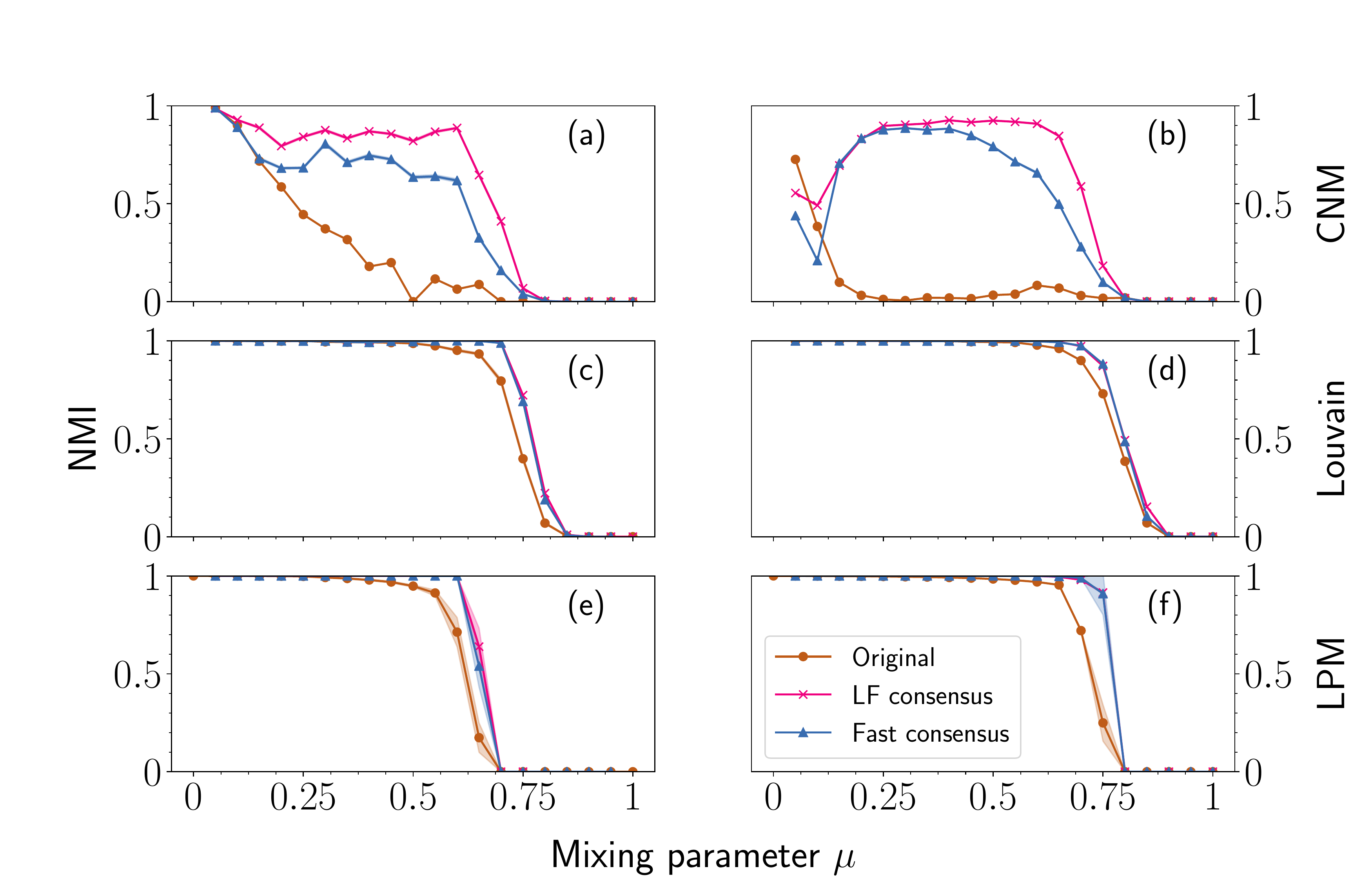}\\
\caption{Performance of fast consensus clustering on LFR benchmark graphs with 1,000 (left) and 10,000 (right) nodes. The algorithms are fast greedy modularity maximization by Clauset, Newman and Moore (CNM, panels (a) and (b)), Louvain (panels (c) and (d)) and the Label Propagation Method (LPM, panels (e) and (f)). The three curves in each plot correspond to the performance of the community detection method (brown), \textit{LF consensus} (magenta) and fast consensus (blue).
The thickness of the lines indicates the standard error of the mean.
The parameters used to generate the LPM benchmark graphs are: degree exponent $\tau_1 = 2$, community size exponent $\tau_2 = 3$, average degree $k_{avg} = 20$, maximum degree $k_{max} = 50$, minimum community size $c_{min} = 10$, maximum community size $c_{max} = 50$ for $n=1,000$ nodes and $c_{max} = 100$ for $n=10,000$ nodes. For each $\mu$-value we generated 20 benchmark configurations, for the LPM and $\mu=0.6$ (1,000 nodes) and $\mu=0.75$ (10,000 nodes) we used $100$ configurations to reduce the error.}
\label{fig:1k10k}
\end{figure*}
In Fig.~\ref{fig:triad_closure} we show that triadic closure helps to improve the accuracy of the results.
The test was carried out by using the Louvain algorithm on LFR benchmark graphs with $10,000$ nodes. The value of the mixing parameter $\mu$ was set to $0.75$, which is in the area where the performance starts degrading (see Fig.~\ref{fig:1k10k}). The other parameters are given in the caption. On the $x$-axis we report the maximum number of triads that can be closed at each iteration, in multiples of the number of links $m$. 
When the number of triads exceeds about $m$, a plateau is reached, yielding a superior accuracy compared to the input partitions and \textit{LF consensus}.

In Ref.~\cite{lancichinetti12} it was shown that consensus clustering leads to more stable partitions compared to the input ones. We want to check if this holds true for our method as well. This analysis is illustrated in Fig.~\ref{fig:stability}, where we computed the average NMI scores between any two input and output partitions, respectively, for the Louvain method on LFR benchmarks. We see that for the consensus partitions the average is very close to one for any value of the mixing parameter $\mu$, including those corresponding to the regime where the clusters are not detected. In contrast, the input partitions become progressively uncorrelated when communities become fuzzy and undetectable. We stress that the result is non-trivial here because we deal with a sparse consensus matrix through the whole calculation, which could introduce a significant amount of noise in the output compared to \textit{LF consensus}, where the consensus matrix is dense and the community structure considerably enhanced. 

In Fig.~\ref{fig:1k10k} we compare the accuracy of our method with \textit{LF consensus}. For each method we identified the value of the threshold parameter $\tau$ that yields the best performance, which is $0.7$ for CNM, $0.2$ for Louvain and $0.8$ for LPM. We used these thresholds systematically.
For Louvain and LPM our method has the same performance as \textit{LF consensus} across all values of $\mu$. For CNM our method does not perform as well as \textit{LF consensus}, but its accuracy is still way above that of the CNM algorithm itself, especially on the larger graphs.

\begin{figure}
 \includegraphics[width=\columnwidth]{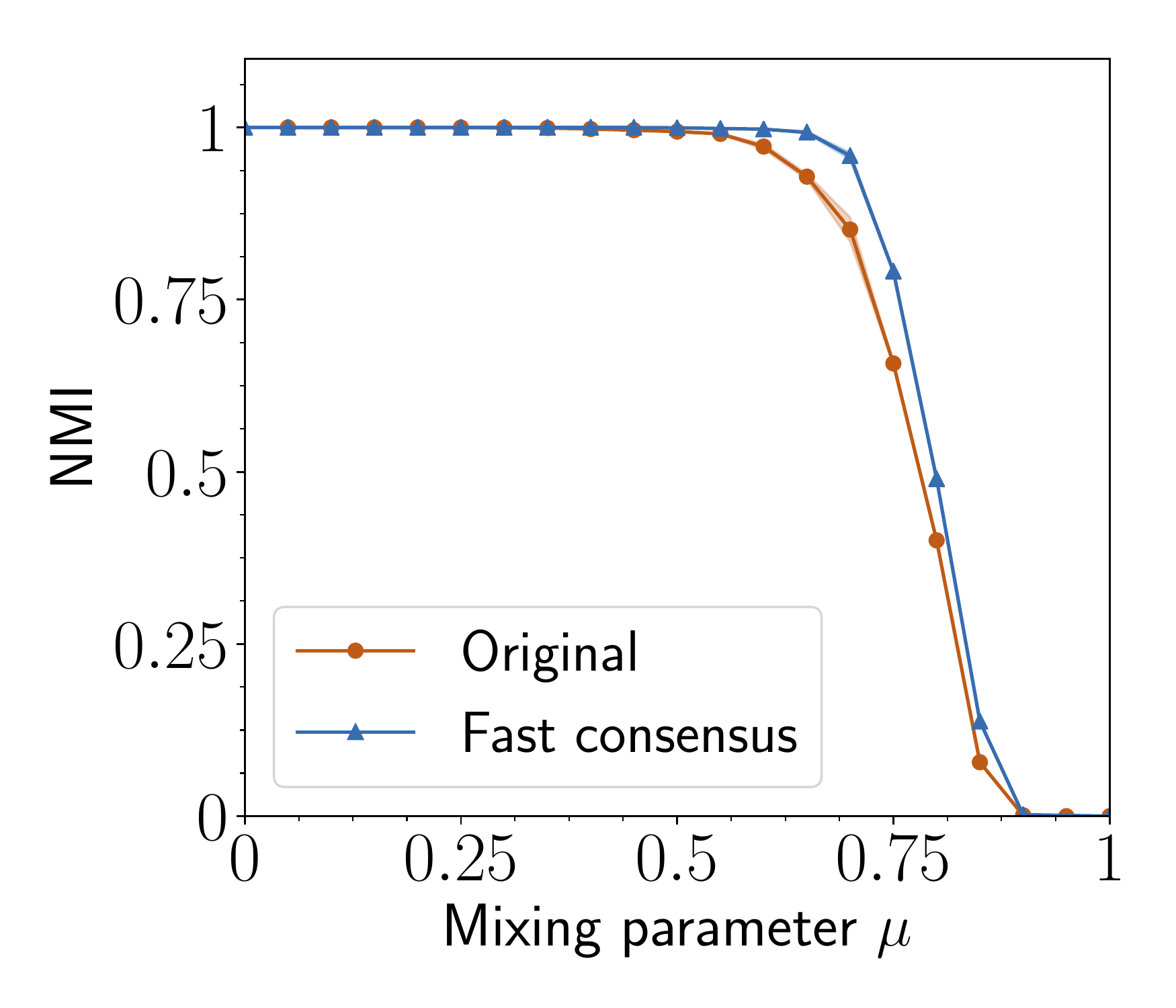}\\
\caption{Fast consensus clustering on large networks.
Comparative analysis of the performance of the Louvain algorithm, with and without consensus clustering, on LFR benchmark graphs with 100,000 nodes. The thickness of the lines indicates the standard error of the mean.
The other parameters to construct the graphs are: degree exponent $\tau_1 = 2$, community size exponent $\tau_2 = 3$, average degree $k_{avg} = 20$, maximum degree $k_{max} = 50$, minimum community size $c_{min} = 10$, maximum community size $c_{max} = 1,000$.}
\label{fig:louvain100k}
\end{figure}

In general, our fast consensus clustering algorithm can be applied to large network sizes, which were hitherto out of reach for \textit{LF consensus} as well as all consensus clustering techniques relying on the calculation of the full consensus matrix. In Fig.~\ref{fig:louvain100k}
we show the performance plot of the Louvain algorithm on LFR graphs with 100,000 nodes. As for the smaller network sizes, our technique outperforms the direct application of the clustering algorithm. The margin is small because the Louvain algorithm (first level communities, see Section~\ref{sec:results}) has already a good performance on the LFR benchmark~\cite{lancichinetti09c}.

\begin{figure}
\centering
 \includegraphics[width=\columnwidth]{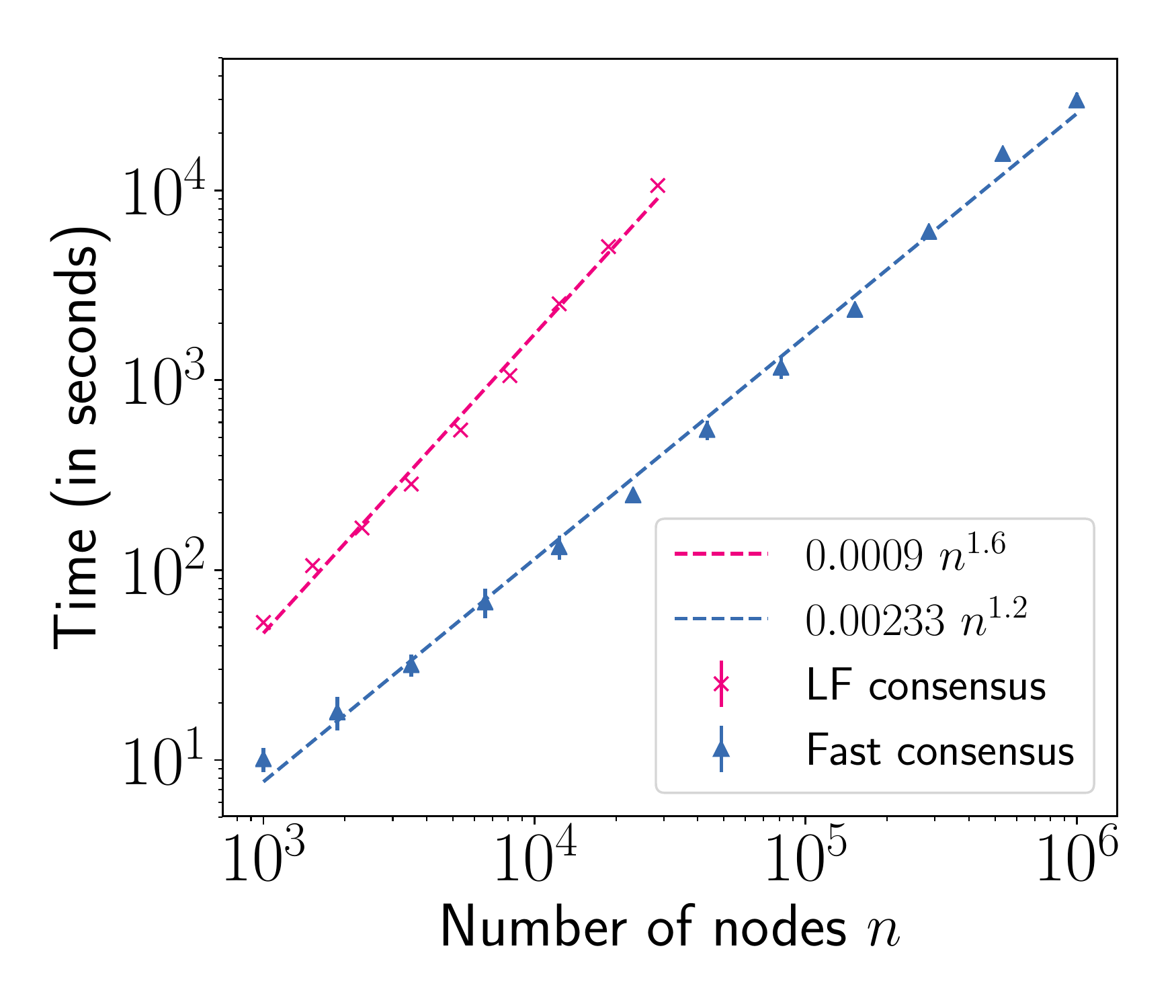}\\
\caption{Time complexity of \textit{LF consensus} and our fast consensus clustering procedure. The clustering algorithm is Louvain, the tests were done on LFR benchmark graphs with the following parameters: degree exponent $\tau_1 = 2$, community size exponent $\tau_2 = 3$, average degree $k_{avg} = 20$, maximum degree $k_{max} = 50$, minimum community size $c_{min} = 10$, maximum community size $c_{max} = n/10$ ($n$ being the number of nodes of the graph). The mixing parameter $\mu$ is set to the value $0.3$. The error bars indicate the standard deviation of the mean.}
\label{fig:scaling}
\end{figure}

Figure~\ref{fig:scaling} compares the time complexity of the consensus clustering procedure presented here with that of Ref.~\cite{lancichinetti12}. Calculations were executed on an iMac with a 3.2 GHz Intel Core i5 processor. We have been able to run our technique on networks with one million nodes in a few hours. Getting consensus partitions from \textit{LF consensus} on networks of this size is impossible because of the high memory and time demands. The complexity of both methods scales as a power of the number of nodes of the network, with exponents $1.6$ (\textit{LF consensus}) and $1.2$ (fast), respectively. Also, the prefactor for the complexity of our method is significantly smaller than for \textit{LF consensus}.
For our method, the complexity of the calculation of the consensus matrix is exactly linear, as we have seen (proportional to $m$, where $m \sim n$ if the network is sparse), and the number of iterations required to reach convergence increases very slowly with the network size. So the final complexity matches that of the clustering algorithm. Indeed, we have verified that the {\tt igraph} implementation of the Louvain algorithm we have used has slightly superlinear complexity, in accord with the exponent $1.2$. For \textit{LF consensus}, instead, the complexity is dominated by the construction of the consensus matrix and it can get actually close to quadratic if the input partitions have large clusters, which severely constrains its applicability. Naturally, if the chosen clustering algorithm has complexity significantly larger than linear, there is  no gain in using a fast consensus approach.

\section{Conclusions}

We have devised an algorithm to implement a fast variant of the consensus clustering routine introduced in Ref.~\cite{lancichinetti12}. This procedure consists in sampling the consensus matrix, instead of computing all its elements, which could lead to a worst-case quadratic space and time complexity. The elements of the consensus matrix which are actually computed are those corresponding to the pairs of neighbors of the network at study, plus at most as many pairs closing triangles with those. The performance remains the same or close to that of the full procedure in all cases we have examined, while the complexity becomes linear, which enables clustering analyses of large networks. 
The idea of sampling the consensus matrix can be easily integrated in other consensus clustering approaches~\cite{bassett13,jeub18,tagarelli17}. We expect that also in those cases the accuracy will remain comparable as in the corresponding original methods. In particular deducting a random baseline from the elements of the consensus matrix, as it is done in the techniques of Refs.~\cite{bassett13,jeub18}, might lead to better results than the simple procedure presented here.

\section{Acknowledgments}
This project was funded by the Deanship of Scientific Research (DSR) at King Abdulaziz University, Jeddah, Saudi Arabia, under Grant No. RG-1439-311-10. The authors, therefore, acknowledge with thanks DSR for technical and financial support.

\appendix
\counterwithin{figure}{section}
\counterwithin{table}{section}

\end{document}